\documentclass[12pt,a4paper,superscriptaddress,preprint,showpacs,aps]{revtex4-1}

\usepackage{amsmath,amssymb}
\usepackage{bm}
\usepackage[dvipdf]{graphicx}
\usepackage{color}
\usepackage{comment}
\usepackage{multirow}

\begin{document}

\title{Complex dynamics of a nonlinear voter model with contrarian agents}
\author{Shoma Tanabe}
\affiliation{Department of Mathematical Informatics,
The University of Tokyo,
7-3-1 Hongo, Bunkyo, Tokyo 113-8656, Japan}
\author{Naoki Masuda}
 \email{masuda@mist.i.u-tokyo.ac.jp}
\affiliation{Department of Mathematical Informatics,
The University of Tokyo,
7-3-1 Hongo, Bunkyo, Tokyo 113-8656, Japan}
\date{December 24, 2013}

\begin{abstract} 
We investigate mean-field dynamics of a nonlinear opinion formation model with congregator and contrarian agents. Each agent assumes one of the two possible states. Congregators imitate the state of other agents with a rate that increases with the number of other agents in the opposite state, as in the linear voter model and nonlinear majority voting models. Contrarians flip the state with a rate that increases with the number of other agents in the same state. The nonlinearity controls the strength of the majority voting and is used as a main bifurcation parameter. We show that the model undergoes
a rich bifurcation scenario comprising the egalitarian equilibrium, two symmetric lopsided equilibria, limit cycle, and coexistence of different types of stable equilibria with intertwining attrative basins.
\end{abstract}


\maketitle

\section*{Lead paragraph}

When multiple competing options are available as in election,
humans influence each other as well as are influenced by external input such as mass media to determine their opinions. Mechanisms of collective opinion formation have been studied with the use of various models including the voter model, Ising model and its variants, and coupled phase oscillators. Here we study dynamical behavior of a nonlinear opinion formation model with some contrarian agents who try to avoid the opinion adopted by others. The remainder of agents are congregator agents that imitate others' opinions with a certain rate, as assumed in many other models.  We mainly investigate the bifurcation scenario of the mean-field dynamics of this agent-based model, where the nonlinearity with which an agent complies with the majority voting is taken as a chief bifurcation parameter.
When the nonlinearity is weak, the model allows egalitarian coexistence of the two competing opinions. At an intermediate level of
nonlinearity, the egalitarian equilibrium is not stable, and a limit cycle appears. With strong nonlinearity, two asymmetric equilibria in which one opinion overwhelms the other opinion are stable. The asymmetric equilibria coexist with the limit cycle, and the structure of the attractive basins is progressively complicated as the nonlinearity is increased.

\section{Introduction}

In society, humans may possess different opinions about the same issue and communicate with each other to update their opinions.
Phenomenology and mechanisms of collective opinion formation have been investigated with the use of various mathematical and agent-based models \cite{Castellano2009RMP,Galam2012book}. A basic assumption employed in most of these models is that individuals tend to mimic others' opinions. Then, all the individuals would eventually select the same opinion, i.e., consensus is reached. Consensus is the expected outcome in, for example, the voter model in finite populations and some infinite networks
\cite{Liggett1985book,Castellano2009RMP,Redner2001book,Krapivsky2010book},
and majority vote models in which individuals tend to adopt the majority's opinion \cite{LiBraunstein2011PRE,Galam2004PhysicaA,Stauffer2004PhysicaA,Borghesi2006PRE,Castellano2009PRE,delaLama2005EPL}.

However, in real society,
pure consensus seems to be an exception rather than a norm \cite{Huckfeldt2004book}. In meanfield populations, idiosyncratic preferences
of individuals \cite{Masuda2010PRE,Masuda2011JSM,Galam1997PhysicaA,Gantert2005AIHP} and
contrarians \cite{Galam2004PhysicaA,Stauffer2004PhysicaA,Borghesi2006PRE,LiBraunstein2011PRE,Kurten2008IJMPB,delaLama2005EPL,Martins2010ACS,Nyczka2013JSP,Masuda2013arxiv}
are among two driving forces to prevent consensus. Mechanisms of coexistence has been also theoretically and numerically investigated
in the context of regional language competition \cite{Castellano2009RMP}.

In the present study, we examine effects of contrarian agents in a nonlinear opinion formation model. Nonlinear opinion formation models with contrarians
have been shown to exhibit phase transition
between a consensus-like phase and an egalitarian
phase
 \cite{Galam2004PhysicaA,Stauffer2004PhysicaA,Borghesi2006PRE,LiBraunstein2011PRE,Kurten2008IJMPB,delaLama2005EPL,Martins2010ACS,Nyczka2013JSP}.
In the consensus-like phase, a majority of individuals takes one of the two opinions.
The egalitarian phase is characterized by a more equal proportions of individuals with different opinions and facilitated by
the existence of contrarian agents or behavior.
A first customary approach to the understanding of such a model is to
look at equilibrium properties, rather than dynamics,
with analysis tools in statistical phyics \cite{Galam2004PhysicaA,Stauffer2004PhysicaA,Borghesi2006PRE,LiBraunstein2011PRE,Kurten2008IJMPB,delaLama2005EPL,Martins2010ACS,Nyczka2013JSP}. A complementary dynamical approach to opinion formation with contrarian agents was recently made with the use of
coupled Kuramoto oscillators with contrarian oscillators
\cite{HongStrogatz2011PRL,HongStrogatz2011PhysRevE,Louzada2012SR}.
However, agents in these studies are implicitly 
assumed to be intrinsic oscillators. Therefore,
it is difficult to link the dynamical results obtained from the oscillator model with contrarians to behavior of the spin-based models such as oscillatory dynamics observed in discrete-time models \cite{Galam2004PhysicaA,Galam2013GEMR}.

We investigate the equilibria and dynamics of a continuous-time nonlinear voter model
with contrarian agents. We extend an agent-based linear voter model
with contrarians proposed in Ref.~\cite{Masuda2013arxiv} by introducing
nonlinear interaction. Agents are not inherent oscillators and do 
assume binary opinions as in the Ising and voter models. 
We numerically examine the mean-field
dynamics of the model in detail.

\section{Model}

We assume that there are two types of agents, congregator and contrarian, in the population.
Each agent possesses either of the two states (i.e., opinions), {\bf 0} and {\bf 1}.
Congregators like other agents (i.e., congregators and contrarians) such that they are eager to posssess the same state as others.
Contrarians oppose other agents such that they try to avoid others' states.
It should be noted that agents do not have to distinguish the type of the partner (i.e., congregator or contrarian) with whom they interact.
We denote by $X$ $(0<X<1)$ and $Y \equiv 1-X$ the fraction of congregators and contrarians in the population, respectively.
We denote by $x$ $(0 \le x \le X)$ and $y$ $(0 \le y \le Y)$ the fraction of congregators and contrarians in state {\bf 1}, respectively.
Therefore, there are a fraction of $X-x$ congregators and $Y-y$ contrarians in state {\bf 0}.

A congregator in state {\bf 0} ({\bf 1}) turns to state {\bf 1} ({\bf 0}) with a rate that increases with the number of agents with state {\bf 1} ({\bf 0}); we assume a well-mixed population.
We introduce nonlinearity to the conversion rate such that the rate equation for the number of congregators in state {\bf 1} is given by
\begin{equation}
 \dot x = r[(X-x)(x+y)^d-x(1-x-y)^d],
 \label{eq:dotc}
\end{equation}
where $r (>0)$ is a parameter, and $d (>0)$ represents the degree of the nonlinearity.
With $d=1$, the conversion rate is proportional to the number of agents in the opposite state, as is the case for the voter model.
There is no nonlinearity in this case.
The model with $d=1$ is equivalent to Model 3 in Ref.~\cite{Masuda2013arxiv}. With $d>1$, the congregators obey a majority rule because $(x+y)^d / (1-x-y)^d > (x+y)/(1-x-y)$ if and only if the density of {\bf 1} agents (i.e., $x+y$) is larger than that of {\bf 0} agents (i.e., $1-x-y$).
In particular, when $d \ge 1$ is integer, Eq.~\eqref{eq:dotc} can be interpreted as a unanimity rule \cite{Lambiotte2007PRE} in which a congregator simultaneously meets randomly selected $d$ agents and flips the state if all the $d$ agents possess the opposite state.
In contrast, Eq.~\eqref{eq:dotc} represents a minority rule when $d<1$.
Similarly, the rate equation for the contrarians is given by
\begin{equation}
 \dot y =  (Y-y)(1-x-y)^d-y(x+y)^d.
 \label{eq:dotl}
\end{equation}
When $r=1$, congregators and contrarians update their states at the same rate.
Equations~\eqref{eq:dotc} and \eqref{eq:dotl} are invariant under the transformation $(x,y) \rightarrow (X-x,Y-y)$, reflecting the symmetry between the two states.
\section{Local stability analysis}
\begin{figure}[htbp]
  \begin{center}
  \includegraphics[width=8cm]{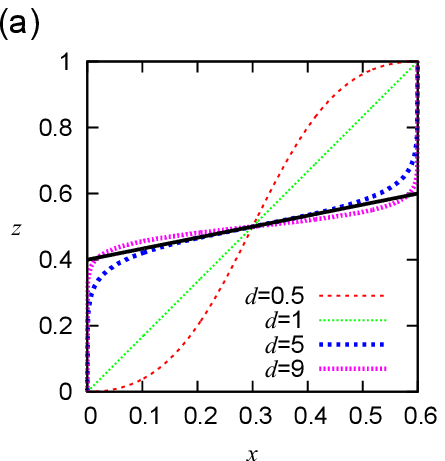}
  \includegraphics[width=8cm]{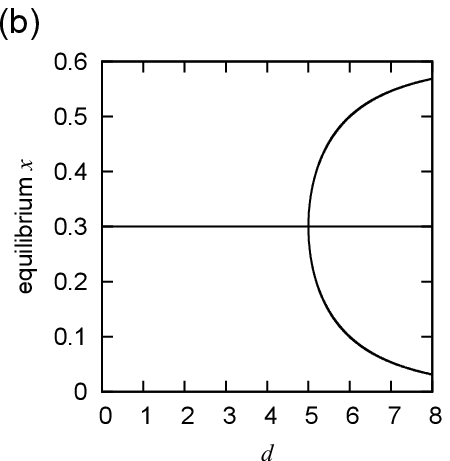}
 \caption{
(a) Nullclines.
The solid line represents Eq.~\eqref{eq:straight} and the four dotted lines represent Eq.~\eqref{eq:curve} with $d=0.5$, $1$, $5$, and $9$.
(b) Bifurcation diagram.
A pitchfork bifurcation occurs at $d=5$ such that three equilibria exist when $d>5$.
We set $X=0.6$ and $Y=0.4$ in both (a) and (b).
}
\label{fig:cy}
\end{center}
\end{figure}

First, we seek the equilibria of the rate equations \eqref{eq:dotc} and \eqref{eq:dotl}.
A linear combination of Eqs.~\eqref{eq:dotc} and \eqref{eq:dotl} under $\dot x=0$ and $\dot y=0$ yields
\begin{equation}
 \frac{x}{X}+\frac{y}{Y}=1.
 \label{eq:straight}
\end{equation}
Equation~\eqref{eq:dotc} under $\dot x=0$ leads to
\begin{equation}
 \displaystyle x+y= \frac {1}{1+\left (\frac{X-x}{x}\right )^{1/d}}.
 \label{eq:curve}
\end{equation}
The intersction of Eqs.~\eqref{eq:straight} and \eqref{eq:curve} on the $x$--$y$ plane gives the equilibria.
We denote by $f(x)$ the right-hand side of Eq.~\eqref{eq:curve}.
Equations~\eqref{eq:straight} and \eqref{eq:curve} are drawn on the $x$--$z$ plane, where $z \equiv x+y$, by the solid line and the dotted lines, respectively, in Fig.~\ref{fig:cy}(a) with $X=0.6$, $Y=0.4$, and different values of $d$.
It should be noted that Eq.~\eqref{eq:straight} does not depend on the $d$ value.
The two lines are invariant under the transformation $(x,z)\rightarrow (X-x,1-z)$.
The two lines cross at $(x,z)=(X/2,1/2)$ for any $d$ such that $(x,y)=(X/2,Y/2)$ is an equilibrium.
We call this equilibrium the egalitarian equilibrium. In the egalitarian equilibrium, the number of congregators in state $\bm 0$ is equal to that of congregators in state $\bm 1$, and the same holds true for the contrarians.

If $d \ge 1$ ($d \le 1$), $f(x)$ is concave (convex) for $0 \le x \le X/2$ and convex (concave) for $X/2 \le x \le X$.
When $(X-Y)/X < f'(X/2)$, there is no other equilibrium.
A pitchfork bifurcation occurs when $(X-Y)/X=f'(X/2)$, i.e.,
\begin{equation}
 d=\frac{1}{X-Y}.
 \label{eq:red}
\end{equation}
Equation~\eqref{eq:red} is satisfied by a positive $d$ value
if $X>Y$.
In Fig.~\ref{fig:cy}(a), Eq.~\eqref{eq:red} is equivalent to $d=5$ because we set $X=0.6$ and $Y=0.4$.
When $(X-Y)/X > f'(X/2)$, there are two other equilibria, whose positions are symmetric about the egalitarian equilibrium ($d=9$ in Fig.~\ref{fig:cy}(a)).
A bifurcation diagram with $d$ being the bifurcation parameter is shown in Fig.~\ref{fig:cy}(b).
It should be noted that the results shown in Fig.~\ref{fig:cy} are independent of the value of $r$.

\begin{figure}[htbp]
\begin{center}
\includegraphics[width=6cm]{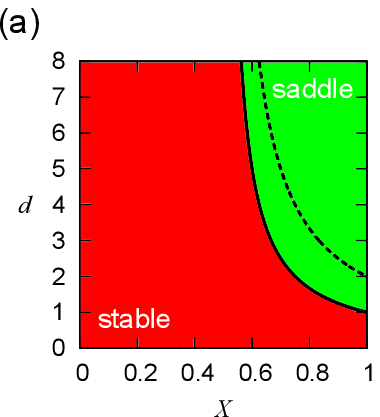}
\includegraphics[width=6cm]{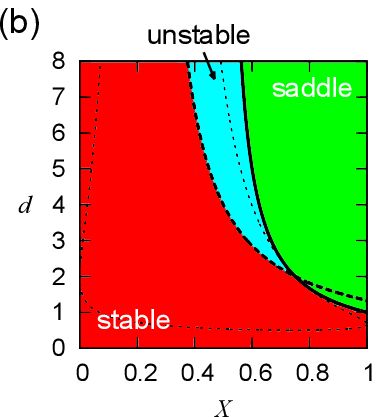}
\includegraphics[width=6cm]{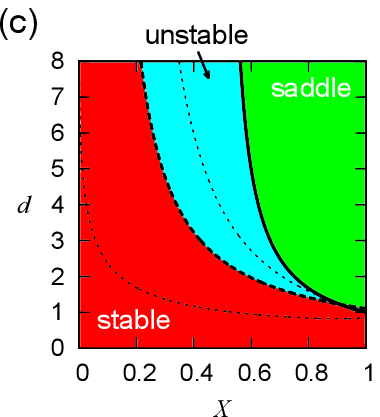}
\end{center}
\caption{Stability of the egalitarian equilibrium.
We set (a) $r=1$, (b) $r=3$, and (c) $r=10$.
The solid lines represent $\det(\tilde J)=0$.
The dashed lines represent ${\rm tr}(\tilde J)=0$.
The dotted lines represent $[{\rm tr}(\tilde J)]^2-4\det(\tilde J)=0$.
In (b) and (c), ${\rm Im}(\lambda_i) \ne 0$ in the regions bounded by the two dotted lines.}
 \label{fig:phase}
\end{figure}

The Jacobian at the egalitarian equilibrium, calculated from Eqs.~\eqref{eq:dotc} and \eqref{eq:dotl}, is given by
\begin{equation}
 J |_{(x,y)=(X/2,Y/2)}=\frac{1}{2^{d-1}} \left[
 \begin{array}{cc}
  r(dX-1) & rdX   \cr
      -dY & -dY-1
 \end{array}
 \right ].
 \label{eq:jacobian}
\end{equation}
For the rescaled Jacobian $\tilde J \equiv 2^{d-1}J |_{(x,y)=(X/2,Y/2)}$, we obtain
\begin{equation}
 \det(\tilde J)=r[1-d(X-Y)]
 \label{eq:red2}
\end{equation}
and
\begin{equation}
 {\rm tr}(\tilde J)=d(rX-Y)-r-1.
  \label{eq:green}
\end{equation}
The egalitarian equilibrium is stable if $\det(\tilde J)>0$ and ${\rm tr}(\tilde J)<0$, saddle if $\det(\tilde J)<0$, and unstable (i.e., both eigenvalues have positive real parts) if $\det(\tilde J)>0$ and ${\rm tr}(\tilde J)>0$.
The parameter values satisfying $\det(\tilde J)=0$ and ${\rm tr}(\tilde J)=0$ for various values of the fraction of congregators (i.e., $X$), $d$, and $r$ are shown by solid and dashed lines in Fig.~\ref{fig:phase}, respectively.
The dotted lines in Fig.~\ref{fig:phase}(b) and \ref{fig:phase}(c) represent the parameter values at which $[{\rm tr}(\tilde J)]^2-4\det(\tilde J)=0$.
The parameters in the regions sandwiched by the two dotted lines yield ${\rm Im}(\tilde \lambda_i) \ne 0$.

Figure~\ref{fig:phase} clarifies the stability of the egalitarian equilibrium.
When $r=1$, the egalitarian equilibrium is never unstable because ${\rm tr}(\tilde J)<0$ whenever $\det(\tilde J)>0$ (Fig.~\ref{fig:phase}(a)).
Because $\det(\tilde J)=0$ is equivalent to Eq.~\eqref{eq:red}, the pitchfork bifurcation occurs on the
solid lines in Fig.~\ref{fig:phase}.
As we increase $X$ or $d$ in Fig.~\ref{fig:phase}(a), the egalitarian equilibrium turns from stable to saddle node via the supercritical pitchfork bifurcation.
The two stable equilibria that are symmetric with respect to the egalitarian equilibrium emerge through the pitchfork bifurcation. We call the two equilibria the lopsided equilibria. The lopsided equilibria correspond to the branches that exist for $d>5$ in Fig.~\ref{fig:cy}(b). In each lopsided equilibrium, one state is dominant in the congregator population, and the other state is dominant in the contrarian population.

When $r=3$ (Fig.~\ref{fig:phase}(b)) and $r=10$ (Fig.~\ref{fig:phase}(c)), the egalitarian equilibrium loses stability via the supercritical pitchfork bifurcation or the supercritical Hopf bifurcation as $X$ or $d$ increases.
For relatively large $X$ and small $d$ values, the egalitarian equilibrium is destabilized via the supercritical pitchfork bifurcation, similar to the case of $r=1$.
For relatively small $X$ and large $d$ values, the egalitarian equilibrium is destabilized via the supercritical Hopf bifurcation such that the egalitarian equilibrium becomes unstable spiral (i.e., two eigenvalues with positive real parts and nonzero imaginary parts) and a limit cycle surrounds the egalitarian equilibrium.
If we further increase $X$ or $d$ to cross the upper dotted line in Fig.~\ref{fig:phase}(b) or \ref{fig:phase}(c), the egalitarian equilibrium turns from unstable spiral to unstable node (i.e., two positive real eigenvalues).
The egalitarian equilibrium turns from unstable node to saddle via the subcritical pitchfork bifurcation as we further increase $X$ or $d$ to cross the solid line in Fig.~\ref{fig:phase}(b) or \ref{fig:phase}(c).
The two lopsided equilibria are unstable upon their appearance.

\section{Global dynamics}
The global dynamics when $r=1$ is implied in Fig.~\ref{fig:phase}(a); there are two phases.
When $X$ or $d$ is relatively small, the egalitarian equilibrium is the unique stable equilibrium.
When $X$ and $d$ are relatively large, the egalitarian equilibrium is saddle and the state space is symmetrically divided into the two attaractive basins of the two lopsided equilibria.

\begin{figure}
\begin{center}
\includegraphics[width=5.3cm]{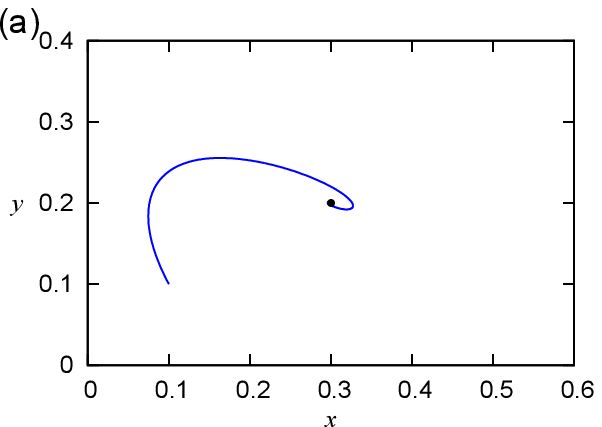}
\includegraphics[width=5.3cm]{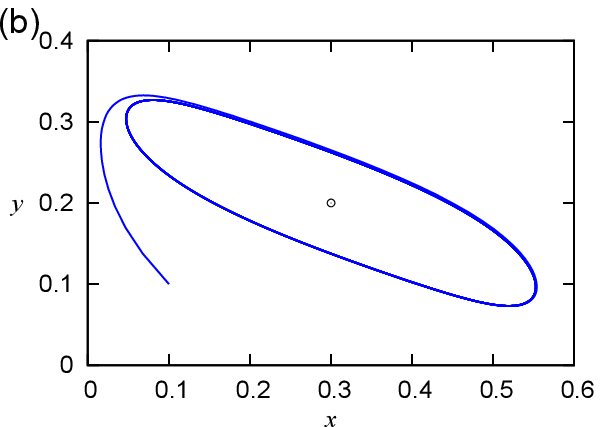}
\includegraphics[width=5.3cm]{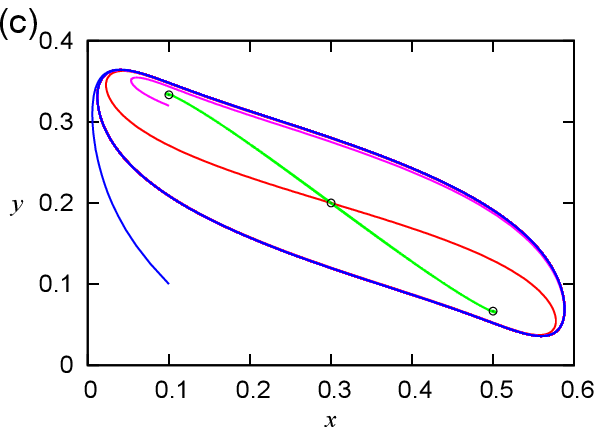}
\includegraphics[width=5.3cm]{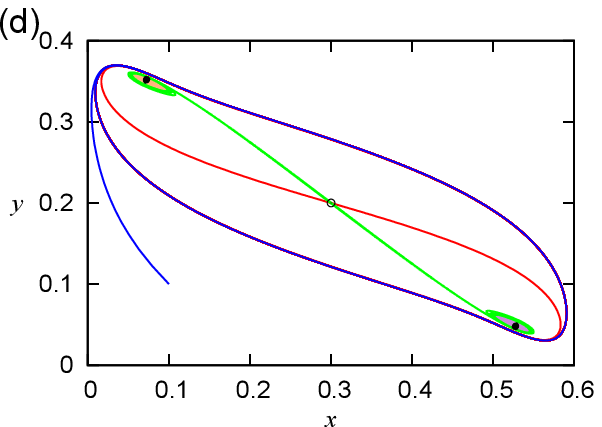}
\includegraphics[width=5.3cm]{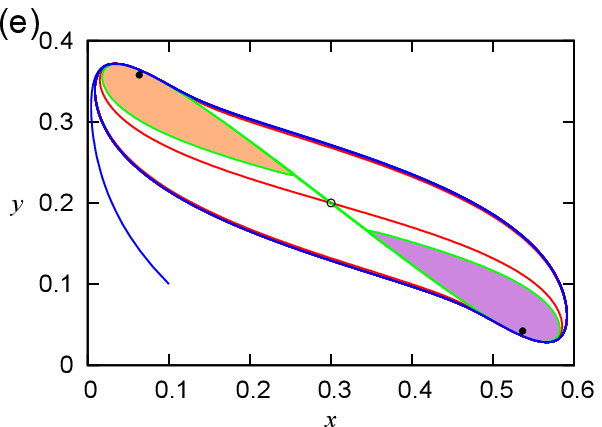}
\includegraphics[width=5.3cm]{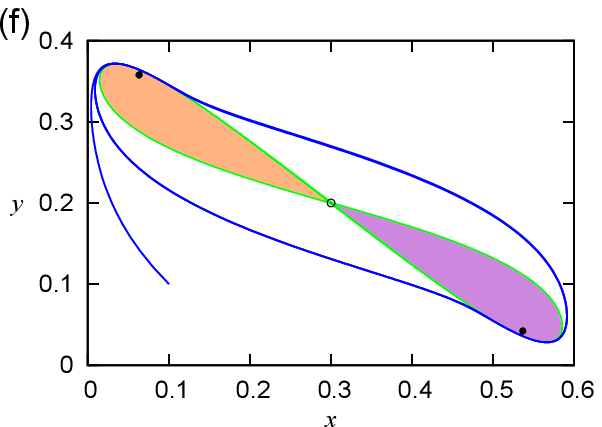}
\includegraphics[width=5.3cm]{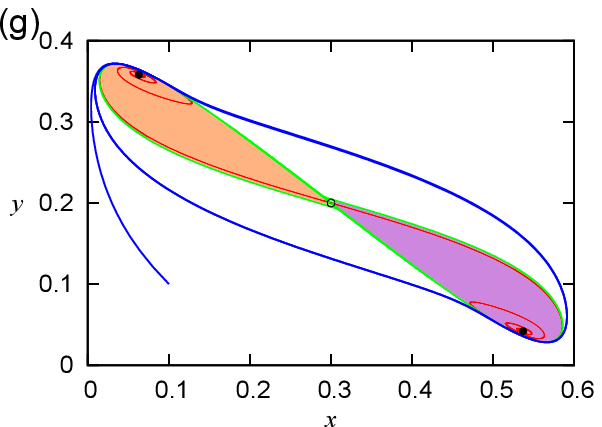}
\includegraphics[width=5.3cm]{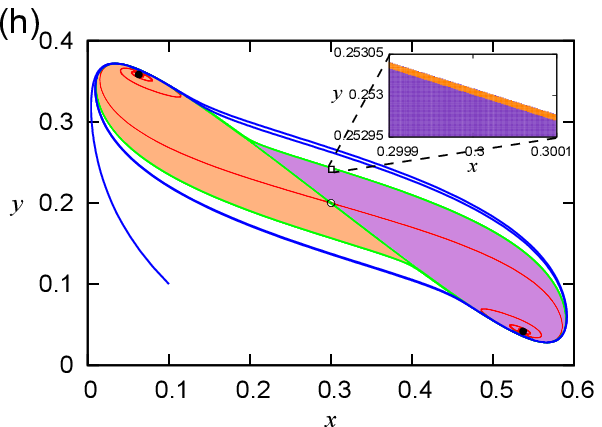}
\includegraphics[width=5.3cm]{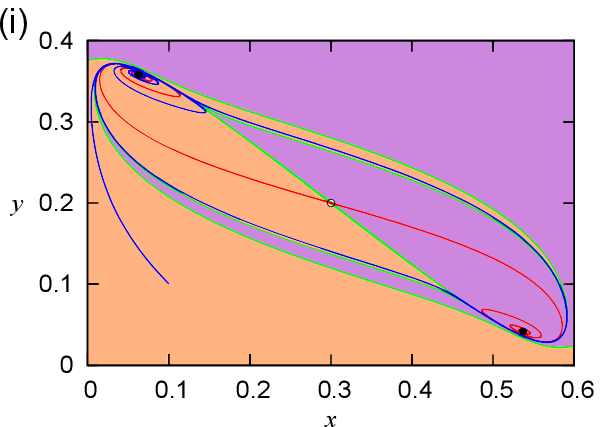}
\end{center}
 \caption{
Dynamics and attractive basins on the $x$--$y$ plane.
We set $X=0.6, Y=0.4$, $r=3$, (a) $d=2.0$, (b) $d=4.0$, (c) $d=6.0$, (d) $d=6.5$, (e) $d=6.71$, (f) $d=6.7196$, (g) $d=6.721$, (h) $d=6.729$, and (i) $d=6.73$.
The filled circles represent the stable equilibrium, i.e., stable node or stable spiral.
The open circles represent the unstable equilibrium, i.e., unstable node or unstable spiral, or saddle.
The green and red lines represent the stable and unstable manifolds of the egalitarian equilibrium, respectively.
The orange and purple shaded regions represent the attractive basins of the two lopsided equilibria.
The blue lines represent the trajectories starting at $(x,y)=(0.1,0.1)$.
The magenta line in (c) represents the trajectory starting at $(x,y)=(0.32,0.1)$.}
 \label{fig:dynamics}
\end{figure}

In the following, we focus on global dynamics of the model when $r=3$.
To understand the dynamics starting from arbitrary initial conditions, we set $X=0.6$ and $Y=0.4$ and numerically investigate the dynamics for various values of $d$ (Fig.~\ref{fig:dynamics}).
Figure~\ref{fig:schematic} schematically summarizes the results shown in Fig.~\ref{fig:dynamics}.

We showed in the previous section that the egalitarian equilibrium experiences the supercritical Hopf bifurcation when $X=0.6$ (Fig.~\ref{fig:phase}(b)).
The Hopf bifurcation occurs at $d \approx 2.86$.
The equilibrium is stable spiral when $d$ is slightly smaller than $d \approx 2.86$ (Fig.~\ref{fig:dynamics}(a); also see Fig.~\ref{fig:schematic}(a) for schematic) and unstable spiral accompanied by a stable limit cycle when $d$ is slightly larger than $d \approx 2.86$ (Figs.~\ref{fig:dynamics}(b) and \ref{fig:schematic}(b)).
The imaginary part of the eigenvalues of the Jacobian matrix at the egalitarian equilibrium vanishes at $d \approx 3.98$ such that the equilibrium turns from unstable spiral to unstable node as $d$ increases (Fig.~\ref{fig:schematic}(c)).
Then, the subcritical pitchfork bifurcation occurs at $d=5$.
When $d$ is slightly larger than five, the egalitarian equilibrium is saddle and two lopsided equilibriua are unstable nodes (Figs.~\ref{fig:dynamics}(c) and \ref{fig:schematic}(d)).

The red lines in Figs.~\ref{fig:dynamics}(c)--\ref{fig:dynamics}(i) represent the unstable manifolds of the egalitarian equilibrium when it is saddle.
The green lines in the same figures represent the stable manifolds of the egalitarian equilibrium.
As $d$ increases, the imaginary part of the eigenvalues of the Jacobian matrix evaluated at the lopsided equilibria becomes nonzero at $d \approx 5.59$ such that the lopsided equilibria turn from unstable nodes to unstable spirals (Fig.~\ref{fig:schematic}(e)).
When the two lopsided equilibria are unstable, regardless of whether they are nodes or spirals, trajectories starting from arbitrary initial conditions eventually tend to the limit cycle (blue and magenta lines in Fig.~\ref{fig:dynamics}(c)).

At $d \approx 6.47$, the lopsided equilibria experience the subcritical Hopf bifurcation such that they become stable and an unstable limit cycle surrounding each lopsided equilibrium appears. 
When $d=6.5$, for example, the region inside each unstable limit cycle is the attractive basin of each lopsided equilibrium (shaded regions in Figs.~\ref{fig:dynamics}(d) and \ref{fig:schematic}(f)).
Any other initial condition either from inside or outside the outer limit cycle is attracted to this limit cycle, except when the initial condition is located on the stable manifold of the egalitarian equilibrium (green lines in Fig.~\ref{fig:dynamics}(d)).
The size of the attractive basin of the lopsided equilibrium increases with $d$ (Fig.~\ref{fig:dynamics}(e)).

A homoclinic bifurcation, in which the stable and unstable manifolds of the egalitarian equilibrium collide, occurs at $d \approx 6.7196$ (Figs.~\ref{fig:dynamics}(f) and \ref{fig:schematic}(g)).
For $d$ values that are a slightly larger $d \approx 6.7196$, the two attractive basins of the lopsided equilibriua infinitely twine with each other (Figs.~\ref{fig:dynamics}(g), \ref{fig:dynamics}(h), and \ref{fig:schematic}(h)).
The boundaries between the intertwined attractive basins are given by the stable manifold of the egalitarian equilibrium (green lines in Figs.~\ref{fig:dynamics}(g) and \ref{fig:dynamics}(h)).
A magnification of the solid square region in Fig.~\ref{fig:dynamics}(h) is shown in the inset.
Trajectories starting from the orange region in the inset once approach the right lopsided equilibrium before they converge to the left lopsided equilibrium.
Trajectories starting from the thin purple region that in fact exists above the orange band in the inset rotates around the two lopsided equilibria once before they converge to the right lopsided equilibrium.

At $d \approx 6.7294$, the stable limit cycle collides with the unstable limit cycle surrounding the two attractive basins of the lopsided equilibria via the saddle--node bifurcation of cycles.
As a result, both the stable and unstable limit cycles disappear for $d>6.7294$.
It should be noted that, different from the conventional saddle--node bifurcation of cycles, the inner unstable limit cycle in the present case in fact consists of an unstable limit cycle inside which the two attractive basins are intermingled with each other.
When $d>6.7294$, initial conditions outside the remnant of the stable limit cycle are attracted to either lopsided equilibrium (blue line in Fig.~\ref{fig:dynamics}(i)) such that the two attractive basins exhaust the entire state space (Figs.~\ref{fig:dynamics}(i) and \ref{fig:schematic}(i)).
The size of the attractive basin for each lopsided equilibrium discontinuously increases at $d \approx 6.7294$.

\section{Conclusions}

We analyzed dynamics of a nonlinear opinion formation model with contrarian agents. 
Our bifurcation analysis revealed that there were three types of stable limit set: the
egalitarian equilibrum (i.e., the two states coexist with the equal fractions), lopsided equilibria (i.e., one state prevails among congregators and the opposite state prevails among the contrarians), and limit cycle. 
The existence of these three types of behavior has been known for opinion formation models with contrarian agents \cite{Galam2004PhysicaA,Stauffer2004PhysicaA,Borghesi2006PRE,LiBraunstein2011PRE,Kurten2008IJMPB,delaLama2005EPL,Martins2010ACS,Nyczka2013JSP,Galam2013GEMR}. Our contribution lies in 
detailed numerical analysis of such a model revealing rich bifurcation scenarios.
The realized behavior depends on the fraction of congregators (i.e., $X$), strength of nonlinear majority voting (i.e., $d$), and the relative speed at which the two types of agents flip the state (i.e., $r$).
The lopsided equilibria and the limit cycle can coexist as stable limit sets
for some parameter values. In this case, the realized behavior
is determined by the initial condition.
The stable limit cycle appears for intermediate values of $d$ under the condition that the congregators update the state faster than the contrarians do
(i.e., $r>1$).
In fact, electoral cyclical patterns have been empirically and theoretically investigated for a long time in social sciences~\cite{miller1973PS,Jeffery2001GP}.

\begin{figure}[htbp]
  \begin{center}
  \includegraphics[width=12cm]{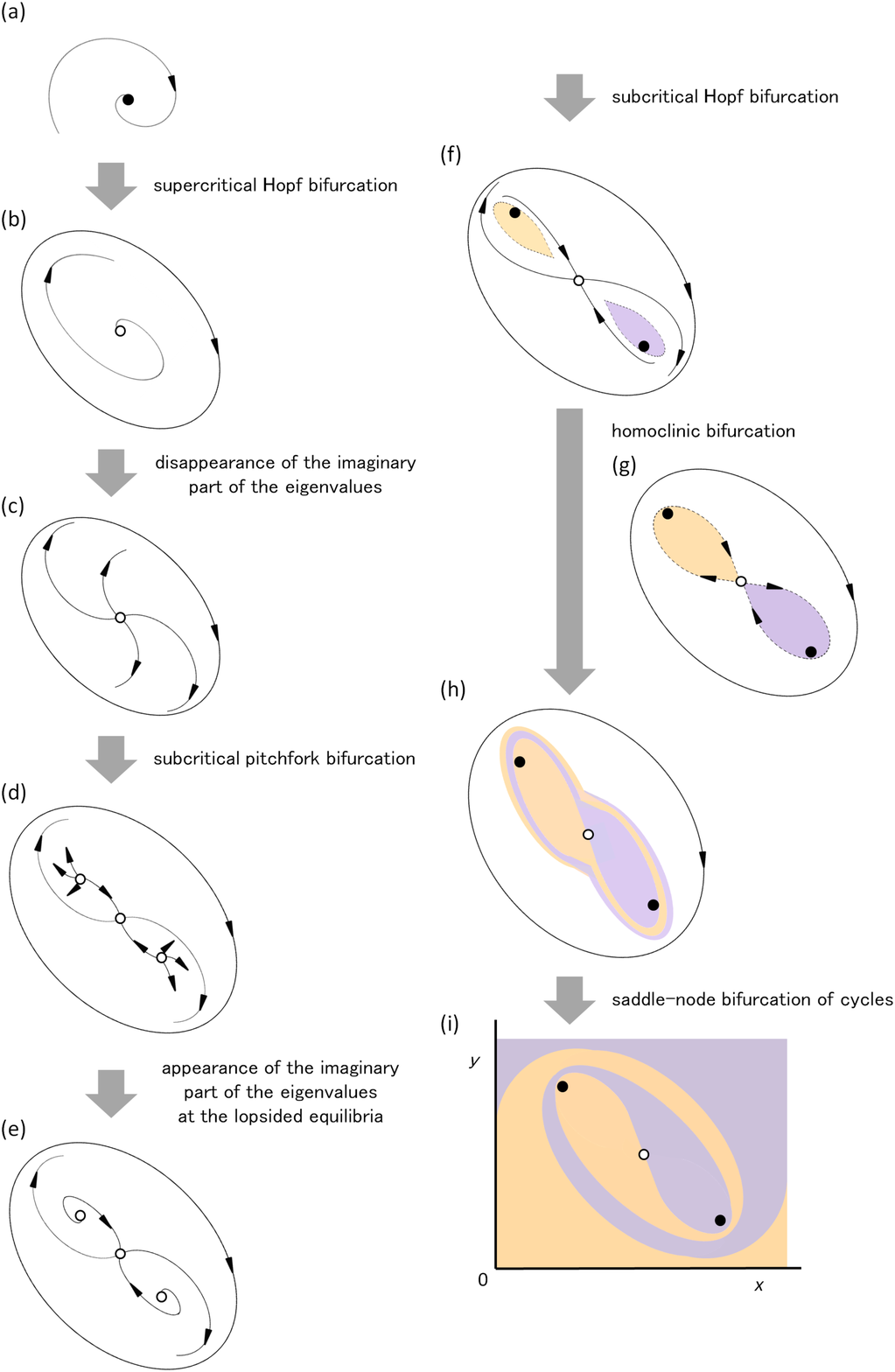}
  \end{center}
 \caption{Schematic illustration of cascades of bifurcations as we increase $d$ with $X=0.6$, $Y=0.4$, and $r=3$.
}
 \label{fig:schematic}
\end{figure}

The same bifurcation scenario as that shown in Fig.~\ref{fig:schematic} is known
for a two-dimensional variant of the so-called Kaldor business cycle model \cite{Bischi2001EE,Agliari2007JEBO}.
However, their model and ours are different.
The Kaldor models represent dynamics of business cycles, and
the two variables in the model considered in Refs.~\cite{Bischi2001EE,Agliari2007JEBO} represent the income level and capital stock.
The two variables interact through \textit{ex ante} investment and saving variables that are functions of the income and capital stock.
In contrast, our model is an opinion formation model
in which agents interact through like-dislike interactions.
The two variables in our model
represent the fractions of agents with a certain opinion in the two subpopulations of agents.
Furthermore, the sequence of bifurcations as shown in 
Fig.~\ref{fig:schematic} is realized by the Kaldor model \cite{Bischi2001EE,Agliari2007JEBO} when the two main parameter values are changed along a nonlinear trajectory (Fig.~1 of \cite{Agliari2007JEBO}).
In our model, the same bifuraction scenario occurs by changing a single bifurcation paremter (i.e., strength of the nonlinear majority voting).


\begin{acknowledgments}
We thank Gouhei Tanaka for discussion.
This work is supported by Grants-in-Aid for Scientific Research (No 23681033)
from MEXT, Japan, the Nakajima Foundation,
and the Aihara Innovative Mathematical
Modelling Project, the Japan Society for the Promotion of Science
(JSPS) through the ``Funding Program for World-Leading Innovative R\&D
on Science and Technology (FIRST Program),'' initiated by the Council
for Science and Technology Policy (CSTP).
\end{acknowledgments}


\begin{thebibliography}{10}

\bibitem{Castellano2009RMP}
C.~Castellano, S.~Fortunato, and V.~Loreto,
\newblock Rev. Mod. Phys. {\bf 81}, 591 (2009).

\bibitem{Galam2012book}
S.~Galam,
\newblock {\em Sociophysics: A Physicist's Modeling of Psycho-political
  Phenomena} (Springer, New York, 2012).

\bibitem{Liggett1985book}
T.~M. Liggett,
\newblock {\em Interacting Particle Systems} (Springer, New York, 1985).

\bibitem{Redner2001book}
S.~Redner,
\newblock {\em A Guide to First-passage Processes} (Cambridge University Press,
  Cambridge, 2001).

\bibitem{Krapivsky2010book}
P.~L. Krapivsky, S.~Redner, and E.~Ben-Naim,
\newblock {\em A Kinetic View of Statistical Physics} (Cambridge University
  Press, Cambridge, 2010).

\bibitem{LiBraunstein2011PRE}
Q.~Li, L.~A. Braunstein, S.~Havlin, and H.~E. Stanley,
\newblock Phys. Rev. E {\bf 84}, 066101 (2011).

\bibitem{Galam2004PhysicaA}
S.~Galam,
\newblock Physica A {\bf 333}, 453 (2004).

\bibitem{Stauffer2004PhysicaA}
D.~Stauffer and J.~S.~S. Martins,
\newblock Physica A {\bf 334}, 558 (2004).

\bibitem{Borghesi2006PRE}
C.~Borghesi and S.~Galam,
\newblock Phys. Rev. E {\bf 73}, 066118 (2006).

\bibitem{Castellano2009PRE}
C.~Castellano, M.~A. Mu{\~n}oz, and R.~Pastor-Satorras,
\newblock Phys. Rev. E {\bf 80}, 041129 (2009).

\bibitem{delaLama2005EPL}
M.~S. de~la Lama, J.~M. L{\'o}pez, and H.~S. Wio,
\newblock Europhys. Lett. {\bf 72}, 851 (2005).

\bibitem{Huckfeldt2004book}
R.~Huckfeldt, P.~E. Johnson, and J.~Sprague,
\newblock {\em Political Disagreement: The Survival of Diverse Opinions within Communication Networks} (Cambridge University Press, Cambridge,
  2004).

\bibitem{Masuda2010PRE}
N.~Masuda, N.~Gibert, and S.~Redner,
\newblock Phys. Rev. E {\bf 82}, 010103 (2010).

\bibitem{Masuda2011JSM}
N.~Masuda and S.~Redner,
\newblock J. Stat. Mech., L02002 (2011).

\bibitem{Galam1997PhysicaA}
S.~Galam,
\newblock Physica A {\bf 238}, 66 (1997).

\bibitem{Gantert2005AIHP}
N.~Gantert, M.~L{\"o}we, and J.~E. Steif,
\newblock Ann. I. H. Poincar{\'e} {\bf 41}, 767 (2005).

\bibitem{Kurten2008IJMPB}
K.~E. K{\"u}rten,
\newblock Int. J. Mod. Phys. B {\bf 22}, 4674 (2008).

\bibitem{Martins2010ACS}
A.~C.~R. Martins and C.~D. Kuba,
\newblock Adv. Comp. Syst. {\bf 13}, 621 (2010).

\bibitem{Nyczka2013JSP}
P.~Nyczka and K.~Sznajd-Weron,
\newblock J. Stat. Phys. {\bf 151}, 174 (2013).

\bibitem{Masuda2013arxiv}
N.~Masuda,
\newblock Phys. Rev. E {\bf 88}, 052803 (2013).

\bibitem{HongStrogatz2011PRL}
H.~Hong and S.~H. Strogatz,
\newblock Phys. Rev. Lett. {\bf 106}, 054102 (2011).

\bibitem{HongStrogatz2011PhysRevE}
H.~Hong and S.~H. Strogatz,
\newblock Phys. Rev. E {\bf 84}, 046202 (2011).

\bibitem{Louzada2012SR}
V.~H.~P. Louzada, N.~A.~M. Ara{\'u}jo, J.~S. Andrade~Jr, and H.~J. Herrmann,
\newblock Sci. Rep. {\bf 2}, 658 (2012).

\bibitem{Galam2013GEMR}
S.~Galam,
\newblock Glob. Econ. Manag. Rev. {\bf 18}, 2 (2013).

\bibitem{Lambiotte2007PRE}
R.~Lambiotte, S.~Thurner, and R.~Hanel,
\newblock Phys. Rev. E {\bf 76}, 046101 (2007).

\bibitem{miller1973PS}
W.~L. Miller and M.~Mackie,
\newblock Polit. Stud. {\bf 21}, 263 (1973).

\bibitem{Jeffery2001GP}
C.~Jeffery and D.~Hough,
\newblock Ger. Polit. {\bf 10}, 73 (2001).

\bibitem{Bischi2001EE}
G.~I. Bischi, R.~Dieci, G.~Rodano, and E.~Saltari,
\newblock J. Evol. Econ. {\bf 11}, 527 (2001).

\bibitem{Agliari2007JEBO}
A.~Agliari, R.~Dieci, and L.~Gardini,
\newblock J. Econ. Behav. Organ. {\bf 62}, 324 (2007).

\end{thebibliography}

\newpage
\clearpage

\clearpage

\clearpage

\clearpage

\end{document}